\documentclass{PoS}
\usepackage{cleveref}
\usepackage{amsmath}
\usepackage{amssymb}
\newcommand{\beq}{\begin{equation}}
\newcommand{\eeq}{\end{equation}}

 \newcommand{\sla}[1]{\displaystyle{\not} #1}
\newcommand{\MSbar}[0]{\overline{\textrm{MS}}}
\newcommand{\pbp}[0]{\ensuremath{\langle \overline{\psi} \psi \rangle}}

\title{Exploring Models for New Physics on the Lattice}

\ShortTitle{Exploring Models for New Physics on the Lattice}

\author{\speaker{Ethan T.~Neil}\\%
       Fermi National Accelerator Laboratory\\
       E-mail: \email{eneil@fnal.gov}}

\abstract{Strongly-coupled gauge theories are an important ingredient in the construction of many extensions of the standard model, particularly for models of electroweak symmetry breaking in which the Higgs boson is a composite object.  There is a large parameter space of such gauge theories with a rich phase structure, particularly in the presence of large numbers of fermionic degrees of freedom.  Lattice simulation provides a non-perturbative way to explore this space of strongly-coupled theories, and to search for interesting dynamical features.  Here I review recent progress in the simulation of such theories, with an eye towards applications to dynamical electroweak symmetry breaking.}

\FullConference{XXIX International Symposium on Lattice Field Theory \\
		 July 10-16, 2011\\
		 Squaw Valley, Lake Tahoe, California}

\begin{document}

\section{Introduction}

Lattice gauge theory has enjoyed considerable success as a non-perturbative method to study and understand quantum chromodynamics (QCD), the only example of a strongly coupled gauge theory observed in nature thus far.  With modern computers and computational techniques, lattice QCD is attaining high levels of precision, with simulation parameters approaching the physical point.  However, the success enjoyed by lattice QCD is no reason for us to restrict the attention of our lattice studies to that theory alone.  QCD is far from unique as an example of an asymptotically free Yang-Mills gauge theory, inhabiting a large space of models, all of which are well-suited to numerical study on the lattice.  Some of these models have properties very different from QCD, lacking phenomena such as confinement and spontaneous breaking of chiral symmetry, and instead showing the restoration of conformal symmetry at low energies.  

These more general Yang-Mills theories are interesting in their own right, particularly in the context of understanding the implications of conformal symmetry in four-dimensional quantum field theory and the AdS/CFT correspondence.  But in the era of the Large Hadron Collider, our primary motivation for the study of theories other than QCD is provided by the search for physics beyond the standard model (BSM).  Strongly-coupled field theories are an essential ingredient for models of dynamical electroweak symmetry breaking, triggered by the spontaneous breaking of chiral symmetry by the new gauge sector.  This can be accomplished either directly with no scalar Higgs field required (as in technicolor \cite{Farhi:1980xs, Hill:2002ap}), or by generating a composite Higgs as a pseudo-Goldstone boson (as in Little Higgs theories \cite{Schmaltz:2005ky,Perelstein:2005ka}).  I will focus on technicolor models as a motivation in this review, since the presence of strongly-coupled physics close to the electroweak scale makes non-perturbative input for such models more urgent.

Such strongly-coupled models can be difficult to work with, as many quantities cannot be calculated using perturbation theory.  Early work on technicolor relied heavily on QCD experiment for non-perturbative input, leading to strong tensions with experiment through the $S$-parameter \cite{Peskin:1991sw} and flavor-changing neutral currents \cite{Eichten:1979ah,Ellis:1980hz}.  But there is no reason to expect more general strongly-coupled theories to lead to the same problems.  Indeed, for many-fermion gauge theories, it has been hypothesized that a theory with approximate scale invariance (colloquially known as a ``walking" theory, since the running gauge coupling evolves relatively slowly) should have the right properties to cure both the flavor-changing neutral current problem \cite{Holdom:1981rm,Yamawaki:1985zg,Appelquist:1986an} and the $S$-parameter \cite{Appelquist:1998xf}.  Lattice simulation provides us with an ideal method to confirm or refute these conjectures non-perturbatively, and to search for other interesting dynamical changes as we move away from QCD and towards the unknown.  (For a review of lattice BSM work with greater focus on these phenomenological issues, see \cite{Fleming:2008gy}.)

The primary focus of this review, and the subject of a recent surge in interest from a number of lattice groups, is on the phase structure of Yang-Mills gauge theories with large numbers of massless fermions.  In section 2, I introduce the concept of the ``conformal window," a range of theories which approach an infrared fixed point at low scales.  Some limited insights from perturbation theory and other analytic arguments are presented.  Section 3 reviews various lattice results for ``QCD-like" theories that are known to lie below the conformal transition.  In section 4, techniques and simulation results for theories which may be inside the conformal window are discussed.  Finally, in section 5 I provide some concluding remarks.

\section{The conformal window \label{sec:CW}}

QCD is our most familiar example of a strongly-coupled Yang-Mills gauge theory, but it is far from unique.  Even restricting our attention to SU$(N_c)$ gauge symmetry groups, there is a vast parameter space available by varying the number of colors $N_c$ and the fermion content of the theory.  Our starting point is the Yang-Mills Lagrangian, which for a theory with $N_f$ massless Dirac fermion species in representation $R$ of the gauge group reads
\beq
\mathcal{L}_{YM} = -\frac{1}{4g^2} F_{\mu \nu}^a F^{a,\mu \nu} + \sum_{i=1}^{N_f} \overline{\psi}_i (i \sla{D}) \psi_i,
\eeq
where $D$ is the usual gauge-covariant derivative.  More generally one may include fermions in multiple representations of the gauge group \cite{Ryttov:2009yw}, but I will consider only the single-representation case here.

At the classical level this Lagrangian is scale invariant, but quantum effects break the conformal symmetry, with the resulting dependence on renormalization energy scale $\mu$ appearing in the renormalized coupling constant $g(\mu)$.  The scale dependence of $g(\mu)$ can be computed order-by-order in perturbation theory, and is usually expressed in terms of the $\beta$-function:
\beq\label{eq3:betaexpand}
\beta(\alpha) \equiv \frac{\partial \alpha}{\partial (\log \mu^2)} = -\beta_0 \alpha^2 - \beta_1 \alpha^3 - \beta_2 \alpha^4 - ...
\eeq
with $\alpha(\mu) \equiv g(\mu)^2 / 4\pi$.  The universal values for the first two coefficients are
\begin{align}\label{eq3:b0}
\beta_0 &= \frac{1}{4\pi} \left(\frac{11}{3} N_c - \frac{4}{3} T(R) N_f \right),\\
\beta_1 &= \frac{1}{(4\pi)^2} \left[\frac{34}{3} N_c^2 - \left(4 C_2(R) + \frac{20}{3} N_c\right) T(R) N_f \right], \label{eq3:b1}
\end{align}
with $T(R)$ and $C_2(R)$ the trace normalization (first Casimir invariant) and quadratic Casimir invariant of representation $R$, respectively.  A list of invariants for some representations of SU$(N)$ are shown in \cref{tab:casimir}; a more exhaustive tabulation of group-theory factors can be found in \cite{Dietrich:2006cm}.  This expansion can of course be continued to higher order, at the price of specifying a renormalization scheme, such as the $\MSbar$ scheme in which the first four coefficients are known \cite{vanRitbergen:1997va}.

\begin{table}[pt]\begin{center}
\begin{tabular}{|c|ccc|}
\hline
Representation&dim$(R)$&$T(R)$&$C_2(R)$\\
\hline
$F$&$N$&$\displaystyle\frac{1}{2}$&$\displaystyle\frac{N^2-1}{2N}$\\[10pt]
$S_2$&$\displaystyle\frac{N(N+1)}{2}$&$\displaystyle\frac{N+2}{2}$&$\displaystyle\frac{(N+2)(N-1)}{N}$\\[10pt]
$A_2$&$\displaystyle\frac{N(N-1)}{2}$&$\displaystyle\frac{N-2}{2}$&$\displaystyle\frac{(N-2)(N+1)}{N}$\\[10pt]
$G$&$N^2-1$&$N$&$N$\\[6pt]
\hline
\end{tabular}
\vspace{1mm}
\caption{Casimir invariants and dimensions of some common representations of $SU(N)$: fundamental ($F$), two-index symmetric ($S_2$), two-index antisymmetric ($A_2$), and adjoint ($G$).\label{tab:casimir}}
\end{center}\end{table}

We can observe several features in the $N_c$-$N_f$ plane based on the two-loop $\beta$-function given above, for some fixed representation $R$.  For any given value of $N_c$, there is a critical number of flavors
\beq
N_f^{AF} = \frac{11N_c}{4T(R)}
\eeq
at which the first coefficient $\beta_0$ changes sign.  Theories with $N_f > N_f^{AF}$ are no longer asymptotically free (the trivial solution $\beta(0) = 0$ is no longer ultraviolet-attractive), and will not be considered further here.  If instead $N_f$ is taken to be large but slightly below $N_f^{AF}$, then for any representation there is a range of values satisfying $\beta_1 < 0$ but $\beta_0 > 0$.  The two-loop $\beta$-function then has a second fixed-point solution \cite{Caswell:1974gg,Banks:1981nn} at non-zero coupling strength:
\beq
\alpha_\star^{(2L)} = -\frac{\beta_0}{\beta_1}.
\eeq
So long as the coupling $\alpha_\star^{(2L)}$ is sufficiently weak, the presence of this infrared-stable fixed point ensures that the theory will be perturbative at all scales, so that the use of the two-loop $\beta$-function is self-consistent.

\begin{figure}
\begin{center}
\includegraphics[width=120mm]{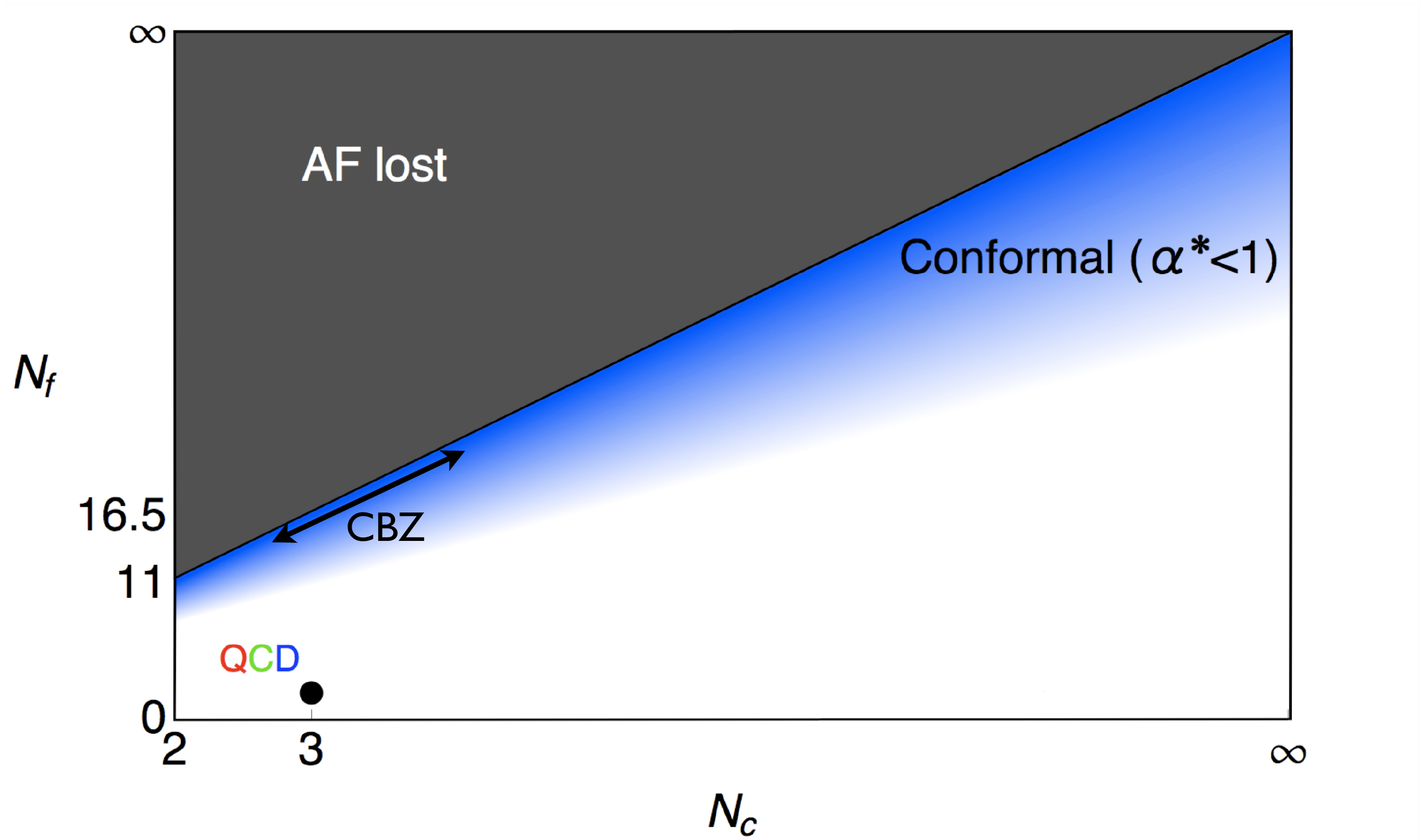}
\caption{Cartoon showing the phase structure in the $N_c-N_f$ plane for SU$(N_c)$ Yang-Mills with fundamental fermions.  The region labelled ``AF lost" contains theories with $\beta_0 < 0$, which are not asymptotically free.  ``CBZ" denotes Caswell-Banks-Zaks theories \cite{Caswell:1974gg,Banks:1981nn}, which flow to very weak infrared fixed points.  The edge of the shaded ``conformal" region, set by the condition $\alpha_\star^{(2L)} < 1$, is intentionally faded to indicate that perturbation theory is becoming unreliable.  Non-perturbative methods must be used to find the true edge of the conformal window.\label{fig:phase-NcNf}}
\end{center}
\end{figure}

As $N_f$ is decreased from $N_f^{AF}$, the perturbative fixed-point coupling increases, and the use of perturbation theory ceases to be valid.  Still, we know that the non-perturbative properties of confinement and spontaneous chiral symmetry breaking should be recovered at small enough $N_f$, in the regime of QCD $(N_f = 2)$ and the quenched $(N_f =0)$ limit.  We define a critical value $N_f^c$ as the point where the infrared behavior of the theory changes from conformal to confining, so that the range of theories
\beq
N_f^c < N_f < \frac{11N_c}{4T(R)}
\eeq
all possess infrared fixed points.  This range is known as the ``conformal window".  Perturbation theory is almost certainly unreliable to describe physics at the infrared fixed point once $N_f$ is close to $N_f^c$, so that study of this transition requires a non-perturbative approach.  A sketch of the behavior outlined here in the $N_c$-$N_f$ plane is shown for the fundamental representation in \cref{fig:phase-NcNf}.

Due to the presence of an infrared fixed point, theories inside the conformal window are not expected to exhibit confinement or spontaneous chiral symmetry breaking.  Conversely, no theory with spontaneously broken chiral symmetry can be in the conformal window, since the infrared fixed point is unstable with respect to the fermion mass $m$, a relevant perturbation.  Another way to state this point is that if the fermions develop a mass $m \neq 0$, then in the infrared limit they will be screened out of the theory \cite{Appelquist:1974tg}, and the coupling will run as in a pure-gauge theory at low scales.

As I have stressed, although perturbation theory is useful for introducing the basic concepts of the conformal window, it ceases to be a useful description near the lower edge.  Lattice simulation provides a fully non-perturbative, albeit numerical, approach to the study of these near-critical theories, and can allow us to constrain $N_f^c$.  Early pioneering studies \cite{Kogut:1985pp,Brown:1992fz,Iwasaki:2003de} were entangled with lattice-artifact phase transitions and unable to draw clear conclusions, but a growing number of lattice groups have undertaken many-fermion simulations over the past few years, and our knowledge of the conformal window in certain classes of theories is rapidly improving.

\begin{figure}[htbp]
\begin{center}
\includegraphics[width=210mm,angle=90]{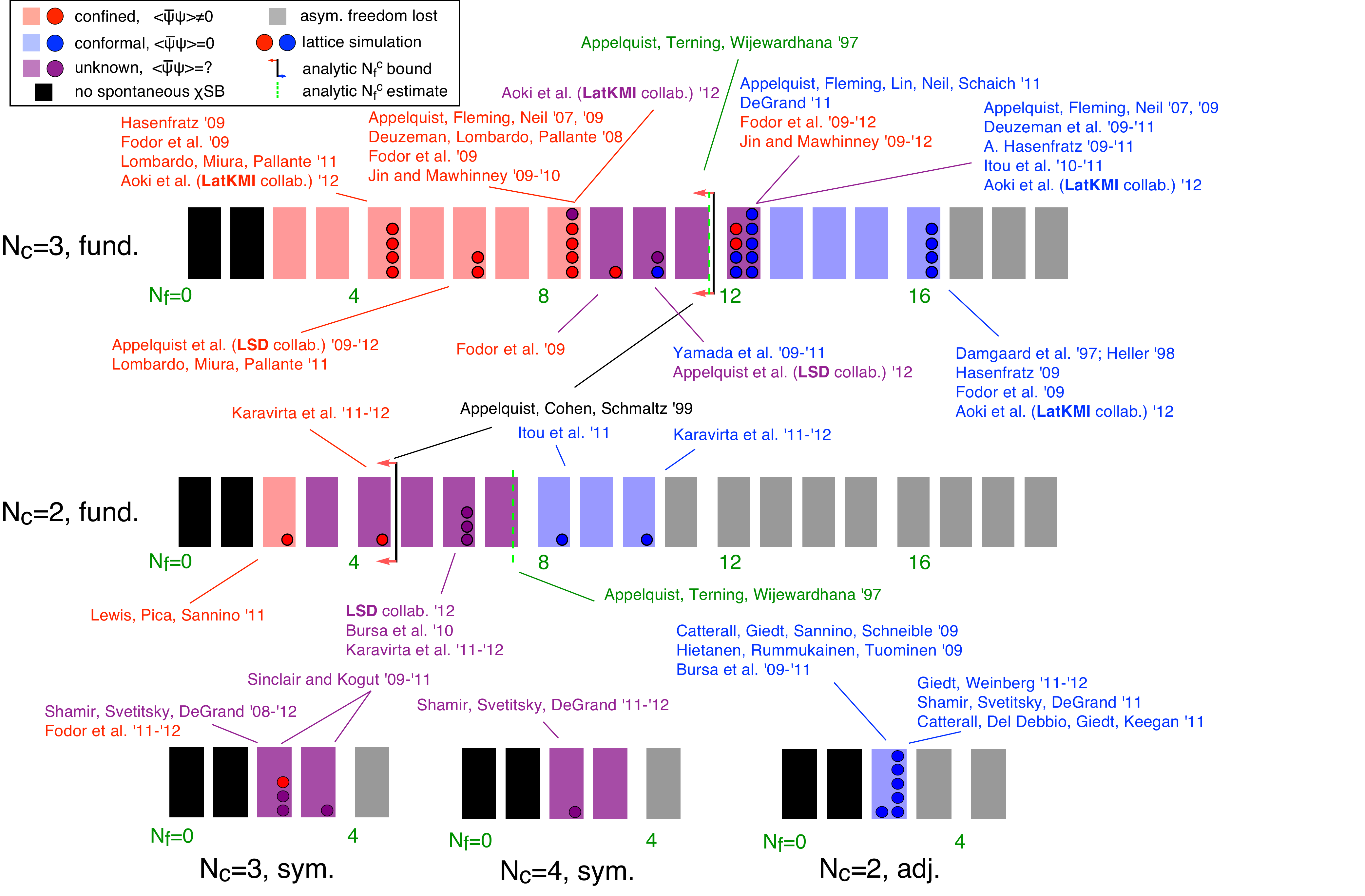}
\caption{Summary plot of recent lattice simulations searching for the conformal window and its properties for various choices of $N_c$ and fermion representation.  Each box represents a particular theory $(N_c, R, N_f)$, and each dot corresponds to the combined results of one lattice group.  \label{fig:CW}}
\end{center}
\end{figure}

In \cref{fig:CW}, I summarize the efforts of lattice groups thus far in attempting to study many-fermion theories.  Each box on the plot represents a particular theory, i.e. a particular choice of the parameters $(N_f, N_c, R)$, and each point represents the combined efforts of one lattice group for that theory.  Simulation results are color-coded based on whether their results indicate confinement (red), conformality (blue), or are inconclusive (purple), based on my reading of the authors' own claims.  Theories themselves are colored similarly, based on my own (subjective) determination of the overall consensus in the field.  Most results thus far are focused on the $N_c = 3, R=F$ case, which contains QCD; these theories allow for the reuse of codes written for QCD with minimal modification.

I have included two analytic approaches to determining $N_f^c$ on this plot.  A direct estimate, denoted by a dashed green line, is obtained by studying the Schwinger-Dyson equation for the fermion propagator \cite{Appelquist:1996dq}.  To make progress in this approach, several assumptions must be made, including a truncation of the set of diagrams contributing to the fermion self-energy known as the ``ladder approximation".  Although some qualitative understanding of the physics behind the conformal transition may be gained with this approach, the error introduced by this truncation is not well controlled at strong coupling, and the perturbative estimates of $N_f^c$ require the further use of the two-loop $\beta$-function.  They are included only for reference.

The other analytic result indicated in this summary plot is a conjectured bound on $N_f^c$ known as the ``thermal inequality" \cite{Appelquist:1999hr}.  The underlying assumption of this inequality is that the number of degrees of freedom should not increase as a theory flows from ultraviolet to infrared; this is an intuitive statement if one considers the RG flow in terms of a ``blocking" operation which tends to reduce the complexity of the system.  The thermal inequality is based on the free energy, which in the free-field limit of a theory with $N_s$ scalars, $N_v$ vectors and $N_f$ Dirac fermions is equal to
\beq
F(T) = -\frac{\pi^2 T^4}{90} \left[ N_s + 2N_v + \frac{7}{8} (4N_f) \right] \equiv -\frac{\pi^2 T^4}{90} f(T).
\eeq
The function $f(T)$ gives a count of the number of light degrees of freedom, so that the conjecture reads
\beq
\lim_{T \rightarrow 0} f(T) \equiv f_{IR} \leq f_{UV} \equiv \lim_{T \rightarrow \infty} f(T). \label{eq:thermal}
\eeq

This inequality can be used to bound $N_f^c$ by applying to a theory in the chirally broken phase.  In the infrared limit, the relevant degrees of freedom for such a theory are the $N_f^2 - 1$ Goldstone bosons associated with the breaking pattern SU$(N_f) \times $ SU$(N_f) \rightarrow $ SU$(N_f)$.  Comparing to the ultraviolet count of free fermions and gauge bosons, we arrive at the result
\beq
N_f \leq 4N_c \left(1 - \frac{16}{81N_c^2}\right)^{1/2},
\eeq
implying that $N_f^c \lesssim 4N_c$.  It should be noted that this relation applies only for $N_c \geq 3$; $N_c = 2$ is a special case.  Because the representations of SU$(2)$ are real or pseudo-real, the bilinear $(\overline{\psi} \gamma^\mu \psi)$ is invariant under an enhanced chiral symmetry group SU$(2N_f)$, rather than the usual SU$(N_f) \times$ SU$(N_f)$.  The breaking pattern preserving the most symmetry is then SU$(2N_f) \rightarrow$ Sp$(2N_f)$, changing the counting for the IR degrees of freedom.  The resulting bound for $N_c = 2$ is then $N_f^c \lesssim 4.7$.

In two dimensions, the conjecture \cref{eq:thermal} is proven for an RG flow connecting two conformal field theories, using Zamolodchikov's $c$-theorem \cite{Zamolodchikov:1986gt} and the direct relation of central charge $c$ to the free energy.  A proof for the generalization of the $c$-theorem to $d=4$ known as the $a$-theorem has recently been presented \cite{Komargodski:2011vj}, but its connection to the free energy is less straightforward, so an $a$-theorem proof does not imply the thermal inequality.  A distinct constraint based on the trace of the energy-momentum tensor in the UV and IR limits does follow from the $a$-theorem, but the resulting bound on $N_f^c$ is very weak, lying above $N_f^{AF}$ \cite{Cardy:1988cwa}.

\section{QCD-like theories}

In using the term ``QCD-like", I am referring specifically to theories with far-infrared properties resembling QCD, namely confinement and spontaneous breaking of chiral symmetry.  These are desirable properties in the context of technicolor or composite Higgs theories, as the broken chiral symmetry is generally used to trigger electroweak symmetry breaking.  (It is also possible to start with a theory inside the conformal window and give masses to some of the fermions, triggering spontaneous symmetry breaking for the massless species, as in e.g. \cite{Luty:2008vs,Galloway:2010bp}.)

Because their infrared properties are similar, lattice simulations of QCD-like theories can rely on mature tools and techniques developed for use in QCD itself.  In particular, the presence of chiral symmetry breaking means that the chiral Lagrangian provides a good effective low-energy description of any QCD-like theory.  Chiral perturbation theory is an essential framework for taking lattice results at the relatively heavy quark/fermion masses which are often required for good control of finite-volume effects, and extrapolating to reach the physical point (for QCD) or the massless limit (in our current context).

SU$(2)$ and SU$(3)$ chiral perturbation theory are familiar to most lattice theorists, and the generalization to arbitrary SU$(N_f)$ is straightforward \cite{Gasser:1986vb}, and has been carried out to next-to-next-to-leading order for several observables \cite{Bijnens:2009qm,Bijnens:2011fm,Bijnens:2011xt}.  Because the higher-order terms contain explicit loop contributions linear in $N_f$, the convergence of chiral perturbation theory is worse with more fermions; the expansion parameter of the theory scales as $N_f M_P^2 / F_P^2$ (up to numerical factors.)  For a fixed lattice spacing $a$, then, we expect that the bare fermion mass should be decreased as $m \propto \sqrt{N_f}$ in order to maintain the stability of the chiral expansion.

The set of theories with $N_c = 3$ and fundamental-representation fermions is a natural choice for the exploration of QCD-like theories, because it contains QCD itself as a reference point.  Furthermore, this choice allows the re-use of existing lattice codes which are heavily optimized for manipulation of SU$(3)$ matrices.  I will focus in this section on theories from $N_f = 2$ to $N_f = 8$, which by  consensus of existing studies are outside the conformal window (as shown in \cref{fig:CW}).

A relatively conservative approach has been taken by the LSD collaboration, who have generated a set of $N_f = 6$ and $N_f = 2$ gauge ensembles, tuned so that confinement-scale observables ($r_0$, $m_\rho$, etc.) are matched in lattice units between the two sets.  They have computed a number of different observables which are particularly interesting in the context of technicolor theories, including the chiral condensate \cite{Appelquist:2009ka} and $S$-parameter \cite{Appelquist:2010xv}.

Some additional details of the latter calculation, which saw a reduction in $S$ for the $N_f = 6$ theory compared to na\"{i}ve scaling predictions, have been released in these proceedings \cite{Schaich:2011qz}.  The issue of finite-volume corrections, a possibly significant source of systematic error in the previous results, is investigated in detail.  By studying the same measurement on smaller volumes ($L/a = 16$), an artificial reduction in $S$ due to finite-volume effects is accompanied with a ``freezing" of the pseudoscalar mass which is not observed on the larger ($L/a=32$) lattices.  Furthermore, a ``pole-dominance" estimate of $S$ which uses sum rules and the properties of the lowest-lying vector and axial-vector states shows a spurious reduction on the larger $L/a = 32$ volume for the $N_f = 2$ estimate, whereas no reduction is seen in the direct computation of $S$.  These results imply that $S$ may be less sensitive to finite-volume effects than other quantities, and that the reduction seen at $N_f = 6$ is physical.

Another quantity recently computed on the $N_f = 2$ and $N_f = 6$ lattices is the $(\Delta I=2)$ $\pi$-$\pi$ scattering length \cite{Appelquist:2012sm}.  In the context of a technicolor theory, where the longitudinal degrees of the $W$ and $Z$ bosons are technipions, this calculation is related to $WW$ scattering, a process which is experimentally accessible at the LHC.  Results for the low-energy constants $\alpha_4$ and $\alpha_5$ in the $N_f = 2$ theory gives values which are likely too small to be probed even with 100 fb$^{-1}$ of LHC data \cite{Eboli:2006wa}.  Extending the analysis to $N_f = 6$ faces further complications, as the appearance of additional low-energy constants in the scattering length makes it impossible to isolate the parameters of interest.  However, an increase in the overall next-to-leading order contribution to the scattering length is observed from $N_f = 2$ to $N_f = 6$. If this qualitative trend continues as $N_f$ increases, then the effect may become observable in the full LHC data set.  Further study will be needed, not only at different $N_f$ but also in different $\pi$-$\pi$ scattering channels.

Moving towards the conformal transition, we turn now to the $N_f = 8$ theory.  Several independent groups using different methods have shown that $N_f = 8$ is outside the window \cite{Appelquist:2007hu,Appelquist:2009ty,Deuzeman:2008pf,Deuzeman:2008sc,Jin:2008rc,Jin:2009mc,Jin:2010vm,Fodor:2009ff,Aoki:2012kr,Miura:2011mc,Miura:2011cy}, so that techniques appropriate for QCD-like theories should still be applicable.  Furthermore, $N_f = 8$ has a direct counterpart in the technicolor literature, the ``one-family" model \cite{Farhi:1980xs} (although because the technifermions in this model carry ordinary color, it is severely constrained by existing LHC searches \cite{Chivukula:2012ug}).  It will be quite interesting to see if the trends indicated at $N_f = 6$ continue as the transition is approached.

However, few simulations have been carried out thus far with a focus on phenomenologically relevant observables in this theory.  The authors of \cite{Fodor:2009ff} have carried out simulations of the Goldstone sector at $N_f = 8$, but are not able to obtain consistent fits to chiral perturbation theory with the available range of masses and volumes.  The LatKMI collaboration has recently begun simulations with the highly improved staggered quark (HISQ) action at every $N_f$ multiple of 4, including $N_f = 8$ \cite{Aoki:2012ep}.  Their results, again focusing on the Goldstone sector, are too preliminary to draw any strong conclusions, but the authors report a clear qualitative difference between $N_f = 4$ and $N_f = 8$.

Another group has attempted to extract the deconfinement temperature $T_c$ \cite{Miura:2011mc,Miura:2011cy}.  They observe a finite-temperature transition at $N_f = 0, 4, 6$, and 8, based on results for the chiral condensate and Polyakov loop.  For the $N_f = 6, 8$ theories, simulations at different $N_t$ are carried out, in order to show the expected scaling behavior with $N_t$ for a real deconfinement transition, $T_c \sim 1/a(\beta_c) N_t$.  (Theories within the conformal window also show deconfinement transitions on the lattice, but they are lattice artifacts and do not persist in the continuum limit; see e.g. \cite{Damgaard:1997ut}.)  The authors go on to compute $T_c / \Lambda_{ref}$, where $\Lambda_{ref}$ is a ``UV reference scale" defined through the two-loop $\beta$-function.  The scale $\Lambda_{ref}$ is dependent on the choice of a reference coupling strength, but for various such choices, the authors observe a decreasing ratio $T_c / \Lambda_{ref}$ as $N_f$ increases.  This result is broadly consistent with analytic estimates of the scaling of $T_c$ \cite{Braun:2009ns}.  The question of the relationship of $T_c$ to other more physical observables, e.g. bound-state masses, remains open.

Finally, I turn to the case of $N_c = 2$ gauge theories (again with fundamental fermions), which have begun to attract more interest recently.  Because the representations of SU$(2)$ are real or pseudo-real, the bilinear $(\overline{\psi} \gamma^\mu \psi)$ is invariant under an enhanced chiral symmetry group SU$(2N_f)$, rather than the usual SU$(N_f) \times$ SU$(N_f)$.  A first study of the spectrum of Goldstone bosons in the $N_c = 2$, $N_f = 2$ theory has been carried out by Lewis, Pica and Sannino \cite{Lewis:2011zb}.  They show the expected number of five Goldstone modes, three pseudoscalar and two scalar, for the symmetry breakdown SU$(4) \rightarrow $Sp$(4)$, and additional bound states which remain heavy in the chiral limit.  SU$(2)$ theories hold particular promise for the application of lattice methods, since the addition of four-fermion interactions and non-zero chemical potential is possible without giving imaginary contributions to the action.

\section{Infrared-conformal theories}

Although substantial progress can be made by studying the properties of QCD-like theories directly, eventually we must turn to the fundamental questions about the conformal window and the nature of the transition at its edge.  Working with theories which are possibly governed by an infrared fixed point introduces many new challenges.  Lattice simulations require the introduction of ultraviolet (lattice spacing) and infrared (box size) cutoff scales, and typically include mass terms for the fermions as well.  Great care must be taken to remove these introduced scales from the problem through continuum, infinite-volume and chiral extrapolations, in order to recover the approximate scale-invariance of the underlying theory.  In some cases, controlled variation of these scales can be used to uncover certain properties of the continuum theory; we will discuss one such example below, using bound-state masses in order to determine the mass anomalous dimension at an infrared fixed point.

As I have done here, the concept of the conformal window is generally first introduced using the perturbative running coupling as a guide.  A natural approach to studying such theories on the lattice is thus through non-perturbative definitions of a running coupling constant, or more broadly, studying the renormalization-group (RG) flow as a function of scale.  This can be a daunting task in theories with many fermions, as the $\beta$-function becomes numerically small, sometimes requiring evolution of the scale over many orders of magnitude in order to see significant RG flow.  All such studies rely on the use of matching techniques to stitch together results obtained on lattices of different physical sizes (it is certainly impossible to change the scale by orders of magnitude while simulating at a fixed UV cutoff.)

The $N_c = 3$, $N_f = 16$ theory provides a good starting point as a theory which is definitely within the conformal window.  One of the earliest lattice calculations directed towards large-$N_f$ theories is \cite{Damgaard:1997ut}, which studied the dependence of this theory on the bare gauge coupling $\beta \equiv 6/g_0^2$.  Despite the fact that $N_f = 16$ is known to have a perturbative infrared fixed point, the authors see clear evidence for a phase transition as $\beta$ is varied, with confinement and chiral symmetry breaking on the strong-coupling side, as indicated by large jumps in the string tension and chiral condensate, respectively.  However, this transition is observed to occur at a fixed $\beta_c$ independent of the simulation volume, indicating that it is associated with a lattice artifact and will not persist in the continuum limit.

In a complementary study \cite{Heller:1997vh}, Heller measured the Schr\"{o}dinger Functional (SF) running coupling in the same theory at bare couplings somewhat below the transition.  The coupling $\bar{g}^2(L)$ was observed to decrease with increasing box size $L$, indicating that the $\beta$-function has the opposite sign relative to QCD.  The simulations at these bare couplings were therefore on the strong-coupling side of the expected IR fixed-point, and therefore disconnected from the continuum limit at zero coupling.  This observation of backwards RG flow is crucial in establishing the existence of a zero of the $\beta$-function using numerical methods with limited precision.

Several different methods for determining the RG flow on the lattice, most of which focus on the extraction of a running coupling constant in a particular scheme, are now in use for a wide range of theories.  I will not go into the details of these various methods here, but refer the interested reader to \cite{DelDebbio:2011rc} for a thorough review.  All of these techniques seek to establish the presence or absence of a zero in some analogue of the $\beta$-function, with the observation of a sign change (i.e. reversal of RG flow) providing the clearest such signal.  The approach of searching for a physical deconfinement temperature $T_c$ also falls into the class of RG-flow methods; the tell-tale signal of a theory outside the conformal window is given by the observation that $T_c$ scales towards zero bare coupling as the temporal extent $N_t$ is increased.  As noted, an example of this method has been applied to the $N_c = 3$, $N_f = 8$ theory, indicating that it lies outside the conformal window \cite{Deuzeman:2008sc,Miura:2011mc}.  A list of references for all of the various RG-flow simulation results is given in \cref{tab:CW}, and the resulting constraints on $N_f^c$ are summarized in \cref{fig:CW}.

Another theory which has attracted considerable interest from lattice groups, and which is now generally agreed to be infrared conformal, is the $N_c = 2$ theory with $N_f = 2$ fermions in the adjoint representation.  Simulations of this theory were first carried out in \cite{Catterall:2007yx} with Wilson fermions, studying the pion decay constant and bound-state masses, but interpretation of the results were complicated by the presence of first-order bulk phase transitions, studied further in \cite{Catterall:2008qk,Catterall:2009sb}.  However, the authors did see some early hints that these scales were vanishing in the massless limit, interpreted as evidence for an infrared-conformal theory.

In the context of this early work, it was not entirely clear how to extrapolate to the massless limit.  For a theory which is inside the conformal window, there is no spontaneous breaking of chiral symmetry, only explicit breaking introduced by the presence of a non-zero fermion mass $m$.  At fixed $m$ this breaking, which also leads to confining behavior in the infrared as the fermions screen out of the theory, leads to a spectrum which resembles a QCD-like theory, with the usual bound states.  However, chiral perturbation theory will not apply in the description of the $m$-dependence of observables, due to the lack of spontaneous chiral symmetry breaking.

\begin{figure}
\begin{center}
\includegraphics[width=140mm]{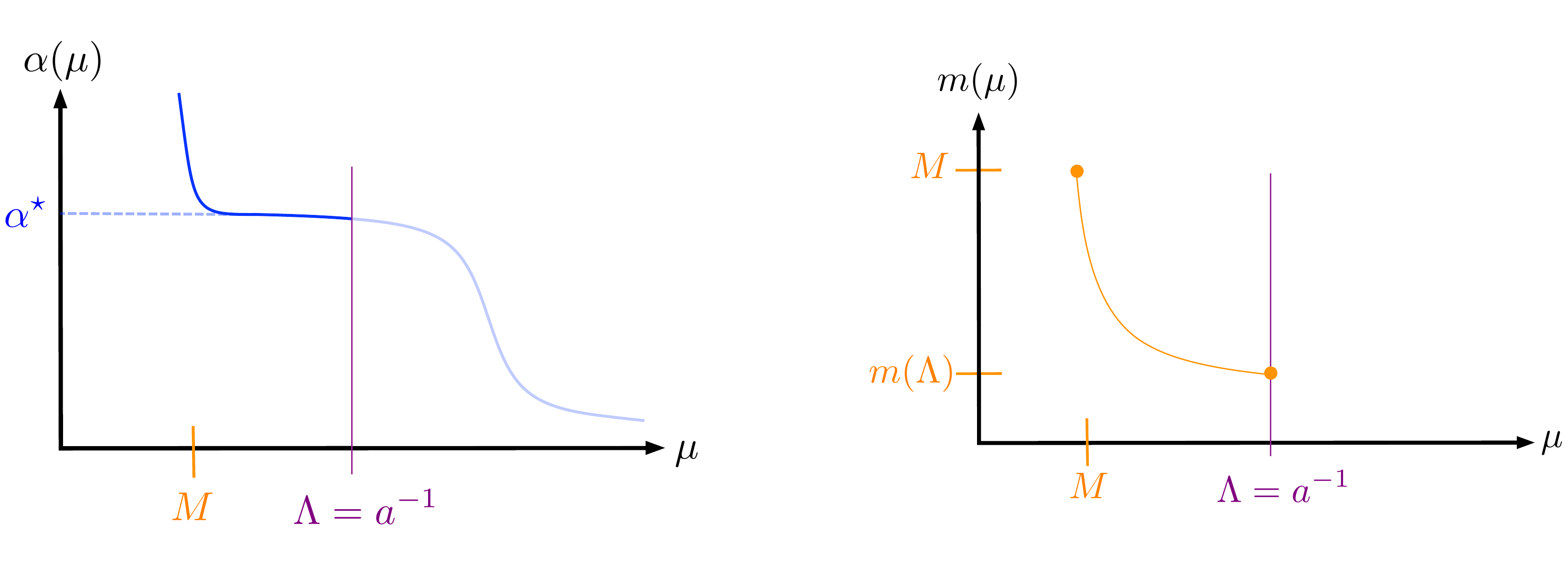}
\caption{Sketch of the renormalization-group evolution in a mass-deformed theory with an IR fixed point.  When a small fermion mass is introduced at UV cutoff $\Lambda$ which is on the ``plateau" $\alpha(\Lambda) \approx \alpha^\star$, the mass will run with the anomalous dimension at the fixed point: $m(\mu) = m(\Lambda) (\Lambda / \mu)^{\gamma^\star}$.  At scale $M$ such that $m(M) = M$, the fermions screen out of the theory and confinement is induced.  The deformed theory has the usual spectrum of bound states, but the mass of any state scales as $m^{1/(1+\gamma^\star)}$ and vanishes in the limit $m \rightarrow 0$. \label{fig:deform}}
\end{center}
\end{figure}

Recent work on the description of bound states in these ``mass-deformed" IR-conformal theories has clarified the situation, providing a framework for extrapolation and giving access to the mass anomalous dimension $\gamma^\star$ as a result \cite{Luty:2008vs,DeGrand:2009hu,DelDebbio:2010ze,DelDebbio:2010jy,Fodor:2011tu,Appelquist:2011dp,DeGrand:2011cu}.  The basic picture is as follows: assume that we are simulating a theory with an infrared fixed point, and that the lattice cutoff $\Lambda = a^{-1}$ has been tuned so that physics on scales below $\Lambda$ is governed by the fixed point (in terms of the gauge coupling, $\alpha(\Lambda) \approx \alpha^\star$).  We now input a small fermion mass, $m = m(\Lambda)$.  The renormalized mass will then run, with its scale dependence determined by the mass anomalous dimension $\gamma^\star$,
\beq
m(\mu) = m(\Lambda) \left( \frac{\Lambda}{\mu}\right)^{\gamma^{\star}}.
\eeq
As $m$ increases, eventually it will reach a scale $M$ such that $m(M) = M$.  At energy scales below $M$, the now heavy fermions will decouple from the theory, and the remaining pure-gauge theory will cause the coupling to run and trigger confinement.  Assuming that this happens over a relatively small range of energy scales, the ``induced" confinement scale is therefore $M$, which is now a function of $m$:
\beq
M = m^{1/(1+\gamma^\star)},
\eeq
where I have now set $a = 1$ to focus on the mass dependence.  A sketch of the overall situation is depicted in \cref{fig:deform}.

As in a QCD-like theory, the masses of bound states will be roughly set by the confinement scale; the main difference is that now the confinement scale itself vanishes as $m \rightarrow 0$, with power-law dependence.  All bound-state masses (including the pions, which are no longer Goldstone modes due to the lack of spontaneous symmetry breaking) will thus scale universally as
\beq
M_X = C_X m^{1/(1+\gamma^\star)}.
\eeq
Decay constants associated with these bound states are expected to scale in the same way \cite{DelDebbio:2010ze,DelDebbio:2010jy}, although simulation results for $F_\pi$ have been observed to be qualitatively different.  The chiral condensate is a special case, which is uniquely sensitive to ultraviolet scales and thus has several additional contributions; the leading quadratic divergence $\pbp \sim m / \Lambda^2$ is present in QCD calculations as well.  The next leading contribution arises from renormalization-group running between the UV and IR scales \cite{Appelquist:2011dp,Patella:2011jr,Fodor:2011tu}:
\beq
\pbp = A_C m + B_C m^{[(3-\gamma^\star)/(1+\gamma^\star)]} + ...
\eeq
Additional terms for this expansion are hypothesized in \cite{Appelquist:2011dp}.  The precise behavior of the chiral condensate as a function of $m$ remains a topic of some debate.  The authors of \cite{Aoki:2012ve} have studied the condensate in this context using the Schwinger-Dyson equation with the ``ladder" truncation, giving some insight into the expected behavior and finite-size scaling.  Recently, Patella has studied the condensate through the spectral density of the Dirac operator, finding that a relatively precise determination of $\gamma^\star$ can be made \cite{Patella:2012da}.

These ``hyperscaling" relations have been applied to spectrum results for the SU$(2)$ $N_f = 2$ adjoint theory, with reasonable success \cite{DelDebbio:2010hx,DelDebbio:2010hu,Kerrane:2010xq,Patella:2010dj,Pica:2010zz,Bursa:2011ru,Appelquist:2011dp}.  Results for the mass anomalous dimension $\gamma^\star$ vary depending on the method used, but are consistently much smaller than 1.  With no spontaneous chiral symmetry breaking and a small $\gamma^\star$, this theory is of limited use for technicolor model-building, but it remains a valuable proving ground for lattice methods designed to study infrared conformality.  An interesting qualitative feature observed by \cite{Patella:2010dj} is that the glueball states in this theory are actually lighter than fermionic bound states, including the pseudoscalar meson.  Such an ordering was predicted as a possible signature of mass-deformed conformal theories simulated on the lattice \cite{Miransky:1998dh}.

The $N_f = 12$ theory with $N_c = 3$ and fundamental fermions has seen a great deal of interest, with a large number of independent groups carrying out simulations; a list of references is given in \cref{tab:CW}.  Although an early study of the SF running coupling indicated a relatively weak infrared fixed point \cite{Appelquist:2007hu,Appelquist:2009ty}, there remains an ongoing debate in the community about whether $N_f = 12$ lies inside the conformal window.  The set of groups studying the RG flow directly have generally found results consistent with an infrared fixed point, while results for the spectrum of bound states are more contentious.  As a particular example, simulation results from Fodor et al. \cite{Fodor:2011tu,Fodor:2012uu} have been fit by the authors using a simple polynomial ansatz and the mass-deformed functional forms shown here, with the conclusion that the IR-conformal fit is disfavored.  However, the authors of \cite{Appelquist:2011dp,DeGrand:2011cu} argue for the opposite conclusion based on slightly different approaches to the mass-deformed fits.  Further and more rigorous development of the framework for dealing with mass-deformed CFTs will be required in order to resolve the disagreement.

All of these discussions are further complicated by the observation of a nontrivial UV fixed point in the theory by \cite{Jin:2012dw,Hasenfratz:2011da}, which can obstruct the ability to extract continuum physics if simulations are carried out too close to the alternate fixed point in bare parameter space.  It is worth noting that all of the simulations carried out at $N_f = 12$ thus far have used staggered fermions.  Certainly staggered fermions are an obvious choice for this theory, as no ``rooting" is needed to thin out the number of continuum flavors as in application to QCD.  However, some unusual results have been observed at strong coupling, in particular the breaking of single-site shift symmetry, manifesting as a difference between the plaquette as measured on adjacent lattice sites \cite{Cheng:2011ic}.  Many of the existing simulation results have been careful to avoid systematic effects induced by these strong-coupling artifacts, but the presence of such complications should be kept in mind and each study evaluated on a case-by-case basis.  It may be worthwhile for lattice groups considering new studies of $N_f = 12$ to consider the use of Wilson or chiral fermions, to lend further clarity to the situation.

Finally, I turn to the 3-color fundamental $N_f = 10$ theory.  There has been much less work here than at $N_f = 12$, but some preliminary results are beginning to appear.  I will discuss the results here in their own right, but note that these simulations cannot be completely separated from the debate at $N_f = 12$.  Any evidence that $N_f = 10$ is infrared conformal necessarily implies that $N_f = 12$ is within the window as well, and vice-versa if $N_f = 12$ is confining and chirally broken.

The first results at $N_f = 10$ were obtained by \cite{Hayakawa:2010yn,Yamada:2010wd}.  Their approach is to compute the Schr\"{o}dinger Functional running coupling, using unimproved Wilson fermions and the plaquette gauge action.  They follow the conventional approach of measuring the SF coupling $g^2(g_0^2, L/a)$ for many bare couplings and volumes, and then use step-scaling to reconstruct a discrete $\beta$-function.  This discrete $\beta$-function is found to be consistent with zero for couplings between roughly $10/3 < g^2 < 9.3$.  More recent results, which have been presented but not yet published, include a measurement of the mass anomalous dimension, pointing to a relatively large value between roughly $0.5 \lesssim \gamma^\star \lesssim 1$ \cite{10f-update}.

Very recently, the LSD collaboration has released spectrum results for the $N_f = 10$ theory \cite{Appelquist:2012nz}.  The results are found to be well-described by a mass-deformed conformal fit, which yields a global best-fit value for the anomalous dimension consistent with $\gamma^\star \sim 1$, in agreement with the Schr\"{o}dinger Functional result.  However, only results for a single lattice volume $32^3 \times 64$ are presently available, and finite-volume corrections (including topological-charge effects) remain an outstanding issue. 

If these preliminary hints of approximate scale invariance with $\gamma^\star \sim 1$ are found to be robust, then $N_f = 10$ will become a theory of great interest.  In the context of technicolor models, it would (if outside the conformal window) be the first proven example of a ``walking technicolor" theory, which could be used as the foundation for building a concrete model without issues with precision constraints.  Even if $N_f = 10$ lies within the conformal window, it can be used to build a technicolor model by giving some of the fermions explicit masses, in order to induce confinement for the remaining fermions \cite{Luty:2008vs,Galloway:2010bp}.  In addition, if it is a theory with a strongly-coupled IR fixed point, the properties of $N_f = 10$ may be of interest from the perspective of AdS/CFT duality.  In particular, a strongly-coupled CFT with a supergravity dual should exhibit a large splitting in the anomalous dimensions of operators with spin greater than two, which correspond to 'stringy' excitations of the dual theory, from the lower-spin operators \cite{Zaffaroni:2000vh}.

\section{Outlook and conclusions}

The task of simulating many-fermion theories provides some novel challenges compared to dealing with QCD.  Even far from the edge of the conformal window, the addition of fermions slows down the running of the gauge coupling, so that much larger changes in physical scale are needed to explore the same range of coupling strengths.  Based on this we might expect that finite-volume corrections become much more problematic in many-fermion gauge theories.  Indeed, the usual rule of thumb $M_\pi L \gtrsim 4$ for QCD simulations to have small finite-volume effects is found to be badly violated in an approximately scale-invariant system \cite{DelDebbio:2011kp}.

In addition to the requirement of large volumes to obtain accurate results, the presence of many fermion species further adds to the cost of HMC evolution.  On top from a trivial scaling with $N_f$ of the number of matrix inversions needed to compute the fermion force, the magnitude of the force also increases with $N_f$, so that the overall cost to maintaining fixed acceptance goes as $N_f^{3/2}$.  All of these issues are compounded by the appearance of factors of $N_f$ in the chiral perturbation theory expansion as discussed, which necessitates moving to lighter fermion masses (and thus even larger volumes.)

The state of the field of many-fermion lattice simulations is similar in many ways to that of QCD simulations many years ago.  Our understanding of these theories is growing rapidly thanks to the combined efforts of many groups, but precise study of near-conformal gauge theories will likely require both increases in computing power and innovations in algorithms and techniques.  In the meantime, answers from the LHC may help us to narrow down the parameter space to a few regions of interest.  Even if the physics behind electroweak symmetry breaking is found to be a strongly-coupled gauge theory that has yet to be simulated, the work carried out thus far on models within and near the conformal window has been crucial in developing our ability to understand and work with such theories.

Finally, although I have not discussed the topic here, it should be noted that there are several important ways that QCD simulations can contribute to searches for BSM physics.  For decay processes which involve mesons and baryons, whether the decay is mediated by the standard model or by new physics, non-perturbative input is required in order to compute the expected rate.  In flavor physics, where BSM theories generically predict a number of effects, uncertainty in QCD contributions remains a significant source of error; for a recent review, see \cite{Laiho:2012ss}, and \cite{Davies:2012qf} in these proceedings.  In some cases, the interaction of BSM physics can require the computation of additional operators that do not appear in standard model processes, including scalar and tensor matrix elements \cite{Bhattacharya:2011qm,Gupta:2012rf} and right-handed currents \cite{Bouchard:2011xj,Bailey:2012wb}.

I would like to thank the organizers for putting together a very successful conference, and for inviting me to review this topic.  I would also like to thank Y. Aoki, C. Bouchard, M. Buchoff, T. DeGrand, G. Fleming, A. Hasenfratz, E. Itou, T. Karavirta, H.-W. Lin, M. Lin, H. Ohki, E. Pallante, A. Patella, D. Schaich, and G. Voronov for providing me with their results prior to my talk, and for many interesting discussions before and during the conference.  Fermilab is operated by Fermi Research Alliance, LLC, under Contract DE-AC02-07CH11359 with the United States Department of Energy.

\begin{table}[htbp]
\begin{center}
\begin{tabular}{cccp{4cm}p{4cm}p{2cm}}
\hline
$R$&$N_c$&$N_f$&Spectrum&RG flow&Other\\
\hline
F&2&2&\cite{Skullerud:2003yc,Muroya:2003jp,Lewis:2011zb}&--&--\\
&&4&\cite{Kogut:2001na}&\cite{Karavirta:2011zg}&--\\
&&6&--&\cite{Bursa:2010xn,Karavirta:2011zg,Karavirta:2012ug}&--\\
&&8&--&\cite{Ohki:2010sr}&--\\
&&10&--&\cite{Karavirta:2011zg,Karavirta:2012ug}&--\\
\hline
F&3&4&\cite{Sui:1998yi,Sui:2001rf}&--&--\\
&&6&\cite{Appelquist:2009ka}&--&\cite{Appelquist:2010xv,Miura:2011mc,Miura:2011cy,Schaich:2011qz,Appelquist:2012sm}\\
&&8&\cite{Jin:2008rc,Jin:2009mc,Fodor:2009ff,Aoki:2012kr}&\cite{Appelquist:2007hu,Appelquist:2009ty,Deuzeman:2008pf,Deuzeman:2008sc,Jin:2010vm}&\cite{Miura:2011mc,Miura:2011cy}\\
&&9&\cite{Fodor:2009ff}&--&--\\
&&10&\cite{Appelquist:2012nz}&\cite{Yamada:2010wd,Hayakawa:2010yn,10f-anomdim}&--\\
&&12&\cite{Jin:2009mc,Jin:2012dw,Fodor:2009ff,Fodor:2011tu,Fodor:2012uu,Appelquist:2011dp,DeGrand:2011cu,Aoki:2012kr}&\cite{Appelquist:2007hu,Appelquist:2009ty,Deuzeman:2009mh,Hasenfratz:2009kz,Hasenfratz:2010fi,Hasenfratz:2011xn,Hasenfratz:2011da,Bilgici:2009nm,Itou:2010we,Aoyama:2011ry,Ogawa:2011ki,Deuzeman:2012pv}&\cite{Cheng:2011ic,Deuzeman:2011pa}\\
&&16&\cite{Damgaard:1997ut}&\cite{Heller:1997vh,Hasenfratz:2009kz,Hasenfratz:2009ea}&\cite{Fodor:2009ff}\\
\hline
G/$S_2$&2&2&\cite{Catterall:2007yx,Catterall:2008qk,Catterall:2009sb,Hietanen:2008vc,Hietanen:2008mr,Bursa:2011ru,DelDebbio:2011kp}&\cite{Hietanen:2009az,DeGrand:2011qd,DeGrand:2011vp,Catterall:2010du,Catterall:2011zf,Catterall:2011ce,Giedt:2011kz,Giedt:2012rj}&--\\
\hline
$S_2$&3&2&\cite{Fodor:2011tw}&\cite{Shamir:2008pb,Svetitsky:2008bw,DeGrand:2008dh,DeGrand:2008kx,DeGrand:2010na,Svetitsky:2010zd,DeGrand:2011vp,DeGrand:2012yq,Sinclair:2009ec,Kogut:2010cz,Sinclair:2010kf,Sinclair:2010be,Kogut:2011ty,Sinclair:2011ie}&--\\
&&3&--&\cite{Sinclair:2010be,Kogut:2011bd}&--\\
\hline
$S_2$&4&2&--&\cite{DeGrand:2011vp,DeGrand:2012qa}&--\\
\hline
\end{tabular}
\caption{Collection of references to simulation results aimed at studying the conformal transition in various classes of SU$(N_c)$ gauge theories.  References are in roughly chronological order, except that subsequent results from a single lattice group are cited consecutively.  The ``other" column encompasses calculations of observables which are not part of the low-lying spectrum and are not used to determine RG flows, for example the $S$-parameter.  Results on $N_f^c$ from these references are summarized in \cref{fig:CW}.  \label{tab:CW}}
\end{center}
\end{table}

\bibliography{bsm-proc-lat11}

\end{document}